\documentclass[a4paper]{jpconf}
\usepackage{graphicx}
\pdfoutput=1
\usepackage{pstricks}
\usepackage{color}
\usepackage{amssymb,amsmath,bbm}
\usepackage{epsf}
\usepackage{epsfig}
\usepackage{afterpage}
\usepackage{longtable}
\usepackage{latexsym,mathrsfs,dsfont}
\usepackage{graphics}
\usepackage{url}
\usepackage{paralist}
\usepackage{bbold}

\newcommand{\gev}{\, {\rm GeV}}

\newcommand{\OmHatEff}{\widehat\Omega_\text{eff}}

\newcommand{\ord}{\mathcal{O}}
\newcommand{\bsi}{B_6^{(1/2)}}
\newcommand{\bei}{B_8^{(3/2)}}

\def\epe{\varepsilon'/\varepsilon}
\newcommand{\be}{\begin{equation}}
\newcommand{\ee}{\end{equation}}

\def\kpn{K^+\rightarrow\pi^+\nu\bar\nu}

\def\klpn{K_{L}\rightarrow\pi^0\nu\bar\nu}

\def \refeq#1{(\ref{#1})}
\newcommand \oL[1]{{\overline{#1}}}
\def \reffig#1{Figure~\ref{#1}}
\def \reftab#1{Table~\ref{#1}}

\DeclareMathOperator{\re}{Re}
\DeclareMathOperator{\im}{Im}

\begin{document}
\title{ The Optimal Strategy for $\epe$ in the SM: 2019}

\author{ Andrzej J Buras}

\address{TUM Institute for Advanced Study, Lichtenbergstr. 2a, D-85748 Garching, Germany}

\ead{aburas@ph.tum.de}

\begin{abstract}
  Following the recent analysis done in collaboration with Jason Aebischer and Christoph Bobeth, I summarize the optimal, in our view,  strategy for the present evaluation of the  ratio $\epe$ in the Standard Model. In particular, I emphasize the   importance of the correct matching of the 
  long-distance and short-distance contributions to $\epe$, which presently is only 
  achieved  by RBC-UKQCD lattice QCD collaboration and by
   the   analytical  Dual QCD approach. An
  important role play also the isospin-breaking  and
  QED effects, which presently are  best known from chiral perturbation theory, albeit still with a significant error. Finally, it is essential to
  include NNLO QCD corrections in order to reduce unphysical renormalization
  scheme and scale dependences present at the NLO level. Here $\mu_c$ in
  $m_c(\mu_c)$ in the case of QCD penguin contributions and $\mu_t$ in $m_t(\mu_t)$ in the case of electroweak penguin contributions play the most important roles. Presently the error on $\epe$ is dominated by the uncertainties  in the QCDP parameter $\bsi$ and the isospin-breaking   parameter $\OmHatEff$.
  We present a table illustrating this.
  \end{abstract}

\section{Introduction}
The present superstars in Kaon physics are in my view
the ratio $\epe$, describing the direct CP violation in $K_L\to\pi\pi$
  decays relative to the indirect one, 
  the $\Delta I=1/2$ rule in $K\to\pi\pi$ decays,
  $K^0-\bar K^0$ mixing with $\varepsilon_K$ and $\Delta M_K$ and of course 
  $\kpn$ and $\klpn$. But there are several decays like $K_L\to\pi^0\ell^+\ell^-$, $K_{S,L}\to\mu^+\mu^-$ and $K_L\to\mu^+e^-$ that
are important for the search for new physics (NP). They were presented in several talks at this conference.

I will concentrate here on $\epe$ which is rather difficult to evaluate.
In order to see the reason for it, let us look at the effective Hamiltonian at the low energy scales. It has the following general structure:
\be\label{BSMH}
{\cal H}_\text{eff}= \sum_i C_i \mathcal{O}_i^\text{SM}+\sum_j C_j^\text{NP} \mathcal{O}_j^\text{NP}\,,\qquad C_i=C_i^\text{SM}+\Delta_i^\text{NP},
\ee
where
\begin{itemize}
\item
$\mathcal{O}_i^\text{SM}$ are local operators present in the Standard Model (SM) 
and $\mathcal{O}_j^\text{NP}$ are new operators having typically new Dirac structures, in particular scalar-scalar and tensor-tensor ones. 
\item
  $C_i$ and   $C_j^\text{NP}$ are the Wilson coefficients of these operators. NP effects
  modify not only the Wilson coefficients of SM operators but also generate
  new operators with non-vanishing $C_j^\text{NP}$.
\end{itemize}

The amplitude for the transition $K\to\pi\pi$  
can now be written as follows
\be
{\cal A}(K\to \pi\pi)= \sum_i C_i\langle \pi\pi| \mathcal{O}_i^\text{SM}|K\rangle
+\sum_j C_j^\text{NP} \langle \pi\pi| \mathcal{O}_j^\text{NP}|K\rangle\,.
\ee

The coefficients $C_i$ and $C_j^\text{NP}$ can  be calculated in the renormalization group (RG) improved perturbation theory.  The status of these calculations is by now very advanced, as reviewed in Ref. \cite{Buras:2011we} and in the talk
by  Maria Cerd{\'a}-Sevilla at this conference. The complete NLO corrections
have been calculated almost 30 years ago \cite{Buras:1991jm,Buras:1992tc, Buras:1992zv, Buras:1993dy, Ciuchini:1992tj, Ciuchini:1993vr}. The dominant NNLO QCD
corrections to electroweak penguin (EWP) contributions have been presented in
\cite{Buras:1999st} and those to QCD penguins (QCDP) should be known soon
\cite{Gorbahn:2004my,Cerda-Sevilla:2016yzo, Cerda-Sevilla:2018hjk}. We will see below that in 
the case of $\epe$ the NNLO QCD contributions play a significant role. On the
whole, the status of present short distance (SD) contributions to $\epe$ is satisfactory.

The evaluation of the hadronic matrix elements is a different story. In
$K\to\pi\pi$ decays, we have presently three approaches  to our disposal:
\begin{itemize}
\item
  {\bf Lattice QCD (LQCD)}. It is a sophisticated numerical method with very demanding calculations lasting many years. Yet, it is based on first principles of QCD
  and eventually in the case of $K\to\pi\pi$ decays and $K^0-\bar K^0$ mixing it
  is expected to give the ultimate results for $\epe$, $\Delta I=1/2$ rule and
  $K^0-\bar K^0$ mixing, both in the SM and beyond it. For $K\to\pi\pi$  only
  results for the SM operators are known and they are not yet satisfactory. The
  ones for BSM $K^0-\bar K^0$ matrix elements are already known with respectable  precision and interesting results have been obtained for long distance
  contributions to $\Delta M_K$ \cite{Bai:2018mdv,Wang:2018csg}.
\item
  {\bf Dual QCD (DQCD)} proposed already in the 1980s \cite{Bardeen:1986vz} and significantly improved
  in this decade \cite{Buras:2014maa}. This approach allows to obtain results for $K\to\pi\pi$ decays
  and $K^0-\bar K^0$ mixing much faster than it is possible with the LQCD so
  that several relevant results have been obtained already in the 1980s and
  confirmed within uncertainties by LQCD in this decade. While not as accurate
  as the expected ultimate LQCD calculations, it allowed already to calculate
  hadronic matrix elements for all BSM operators entering $K\to\pi\pi$ decays
  \cite{Aebischer:2018rrz}
  and $K^0-\bar K^0$ mixing \cite{Buras:2018lgu}.
 The latter paper  allowed to get the insight into the
  QCD dynamics at low energy scales which is not possible using a purely
  numerical method like LQCD. For a recent review see Ref. \cite{Buras:2018hze}. More about it below.
\item
  {\bf Chiral Perturbation Theory (ChPT)} developed since 1978 \cite{Weinberg:1978kz,Gasser:1983yg,Gasser:1984gg,Ecker:1988te,Ecker:1994gg,Pich:1995bw,Cirigliano:2011ny} and discussed in several talks at this conference, in
  particular by Antonio Rodriguez S{\'a}nchez \cite{Cirigliano:2019zjv} 
  and Toni Pich \cite{Cirigliano:2019ani} in the contex of $\epe$. It is based on global symmetries of QCD with the QCD dynamics
  parametrized by low-energy constants $L_i$ that enter the counter terms
  in meson loop calculations. $L_i$ can be extracted
  from the data or calculated by LQCD but to this end the large $N$ limit
  has to be taken. In the case of non-leptonic transitions this
  implies serious  difficulties in matching
  long distance (LD) and short distance (SD) contributions in this framework.
  The point is that in the large $N$ limit only factorizable contributions
  in hadronic matrix elements are present, whereas  the dominant QCD dynamics in
  Wilson coefficients is given by non-factorizable contributions. This problem is
  absent in LQCD and DQCD as we will discuss below. 
  Therefore,
  while the ChPT approach is very suitable for leptonic and semi-leptonic Kaon
  decays, it can only provide partial information on $\epe$ and $\Delta I=1/2$
  rule in the form of isospin breaking effects and final state interactions (FSI). Yet, in the case of isospin breaking contributions to $\epe$ the difficulties  in the matching in question imply a rather significant error as we will see
  below.
  \end{itemize}
  
This writing is arranged as follows. In Section~\ref{sec:2} we will briefly
describe the DQCD approach. In Section~\ref{sec:3} we will present, in our view,
the optimal strategy for the calculation of $\epe$
in the SM as of 2019, illustrating in particular the importance of NNLO
QCD effects and isospin breaking corrections in the evaluation of this ratio.
This presentation is fully based on the recent analysis of $\epe$ in
collaboration with Jason Aebischer and Christoph Bobeth
\cite{Aebischer:2019mtr}. A brief outlook in Section~\ref{sec:5} ends this presentation.

\section{ Grand View on the Dual QCD Approach}\label{sec:2}
This analytic approach to $K\to\pi\pi$ decays and $K^0-\bar K^0$ mixing in Refs.  \cite{Bardeen:1986vz,Bardeen:1987vg,Buras:2014maa} is based on 
the ideas of 't Hooft and Witten who studied QCD with a
 large number $N$ of colours. In this limit  QCD is dual to a  theory of weakly interacting mesons with the coupling $\ord(1/N)$ and in particular in the strict large $N$ limit it becomes a free theory 
of mesons, simplifying the calculations significantly. With  non-interacting mesons the factorization of matrix elements 
of four-quark operators  into matrix elements of quark currents and quark 
densities,  used adhoc in the 1970s and early 1980s, is automatic and can be considered as a property of QCD in this limit \cite{Buras:1985xv}. But the factorization cannot be the whole 
story as the most important QCD effects related to asymptotic freedom 
are related to non-factorizable contributions generated by exchanges of gluons.
In DQCD this role is played by meson loops that represent 
dominant non-factorizable contributions at the very low energy scales. Calculating these loops with a momentum cut-off $\Lambda$ one finds
then  that the factorization in question 
does not take place at values of $\mu \ge 1\gev$ at which Wilson coefficients 
are calculated, but  rather at very low  momentum transfer between colour-singlet currents or densities.  

Thus, even if  in the large $N$ limit the hadronic matrix elements factorize
and can easily be calculated, 
in order to combine them with the Wilson coefficients, loops in the meson 
theory have to be calculated. In contrast to chiral perturbation theory, in 
DQCD a physical cut-off $\Lambda$ is used in the integration over loop momenta.
 As discussed in detail in Refs. \cite{Bardeen:1986vz,Buras:2014maa} this allows 
to achieve  a much better matching with short distance contributions than it
is possible in ChPT, which uses dimensional regularization.  The cut-off $\Lambda$ is typically chosen around $0.7\gev$ when only pseudoscalar mesons are exchanged in the
loops \cite{Bardeen:1986vz} and can be increased up to $0.9\gev$ when contributions from lowest-lying vector mesons are taken into account as done in Ref.  \cite{Buras:2014maa}. These calculations are done in a momentum scheme, but as 
demonstrated in \cite{Buras:2014maa}, they can be matched to the commonly 
used naive dimensional regularization (NDR) scheme. Once this is done it is justified to set $\Lambda\approx \mu$.
We ask sceptical readers to study a detailed exposition of DQCD in Ref. \cite{Buras:2014maa}, where also the differences from the usual ChPT calculations are emphasized.

The application of DQCD to weak decays consists in any  NP model of the following 
steps:

{\bf Step 1:} At $\Lambda_\text{NP}$ one integrates out the heavy degrees of freedom and performs the RG evolution including Yukawa couplings and all gauge interactions present in the SM down to the electroweak scale. This evolution involves in addition to SM 
operators also beyon the SM (BSM) operators. This is the Standard Model effective field theory (SMEFT).

{\bf Step 2:} At the electroweak scale $W$, $Z$, top quark and the Higgs are integrated out and the SMEFT is matched onto the effective field theory with only SM quarks except the top-quark, the photon and the gluons. Subsequently QCD and QED evolution is performed down to scales $\ord(1\gev)$.

{\bf Step 3:} Around scales $\ord(1\gev)$ the matching to the theory of mesons
is performed and the so-called {\em meson evolution} to the factorization scale is performed.

{\bf Step 4:} The matrix elements of all operators are calculated in the 
large $N$ limit, that is using factorization of matrix elements into products of currents or densities.

We do not claim that these are all QCD effects responsible for non-leptonic 
transitions, but these evolutions based entirely on non-factorizable QCD
effects, both at short distance and long distance scales, appear to be the main 
bulk of QCD dynamics responsible for the $\Delta I=1/2$ rule, $\epe$ and $K^0-\bar K^0$ mixing. Past successes of this approach have been reviewed in Refs.
\cite{Buras:2018hze,Buras:2018wmb}. They are related
in particular the non-perturbative parameter $\hat B_K$ in $K^0-\bar K^0$
mixing and  $\Delta I=1/2$ rule \cite{Buras:2014maa}.
In fact DQCD allowed for the first time to identify already in 1986
the dominant mechanism behind this rule \cite{Bardeen:1986vz}.

In 2018 a significant progress   towards the general search for  NP in $\epe$ with the help of DQCD 
 has been made:
\begin{itemize}
\item
The first to date calculations of the $K\to\pi\pi$ matrix elements of 
the chromo-magnetic dipole operators \cite{Buras:2018evv} that
are compatible with the LQCD results for $K\to\pi$ matrix elements
of these operators obtained earlier in \cite{Constantinou:2017sgv}.
\item
The calculation of $K\to\pi\pi$ matrix elements of {\em all} four-quark BSM operators,
including scalar and tensor operators, by DQCD \cite{Aebischer:2018rrz}.
\item
The derivation of a master formula for $\epe$ \cite{Aebischer:2018quc}, which
can be applied to any theory beyond the SM in which the
Wilson coefficients of all contributing operators have been calculated at the
electroweak scale. The relevant hadronic matrix elements of BSM operators used 
in this formula are
from the DQCD, as lattice QCD did not calculate them yet, 
and the SM ones from LQCD.
\item
This allowed to perform the first to date  model-independent anatomy of the ratio $\epe$
in the context of  the $\Delta S = 1$ effective theory with operators invariant
under QCD and QED and in the context of the SMEFT with the operators invariant under the full SM gauge group \cite{Aebischer:2018csl}.
\item
  Finally the insight from DQCD \cite{Buras:2018lgu} into the values of BSM $K^0-\bar K^0$ elements obtained by LQCD  made sure that the meson evolution is
  hidden in lattice calculations.
\end{itemize}

The main messages from these papers are as follows:
\begin{itemize}
\item
  The inclusion of the meson evolution in the phenomenology of any non-leptonic
  transition like $K^0-\bar K^0$ mixing and $K\to\pi\pi$ decays with $\epe$ and
  the $\Delta I=1/2$ rule is mandatory!
\item
  Meson evolution is hidden in LQCD results, but among analytic approaches only
  DQCD takes this important QCD dynamics into account. Whether meson evolution
  is present in the low energy constants $L_i$ of ChPT is an interesting question, still to be answered.
\item
  Most importantly, the meson evolution  
turns out to have the pattern of operator mixing, both for SM and BSM operators,
to agree with the one found perturbatively at short distance scales. This allows for a satisfactory, even if approximate, matching between Wilson coefficients and hadronic matrix elements.
\end{itemize}

In summary DQCD turns out to be an efficient approximate method for obtaining results for non-leptonic decays, years and even decades, before useful results from numerically sophisticated and demanding lattice calculations could be obtained.

\boldmath
\section{$\epe$ in the SM}\label{sec:3}
\unboldmath
\subsection{Preliminaries}

The situation of $\epe$ in the SM after the International Conference on Kaon Physics 2019 can be briefly
summarized as follows:
\begin{itemize}
\item The analysis of $\epe$ by the RBC-UKQCD LQCD collaboration
  based on their 2015 results for $K\to \pi\pi$ matrix elements \cite{Bai:2015nea,
    Blum:2015ywa}, as well as the analyses performed in \cite{Buras:2015yba,
    Kitahara:2016nld} that are based on the same matrix elements but also
  include isospin breaking effects, found $\epe$ in the ballpark of
  $(1-2) \times 10^{-4}$. This is by one order of magnitude below the
  experimental world average from NA48 \cite{Batley:2002gn} and
  KTeV \cite{AlaviHarati:2002ye, Worcester:2009qt} collaborations,
  \be
    \label{EXP}
    \boxed{(\epe)_\text{exp} 
    = (16.6 \pm 2.3) \times 10^{-4}\,.}
  \ee
  However, with an error in the ballpark of $5 \times 10^{-4}$ obtained in
  these analyses, one can talk about an $\epe$ anomaly of at most~$3\,\sigma$.
  The RBC-UKQCD collaboration is expected to present soon new
  values of the $K\to\pi\pi$ hadronic matrix elements.  Not only statistical
  errors have been significantly decreased, but also a better agreement with the
  experimental values of $\pi\pi$-strong-interaction phases $\delta_{0,2}$ has
  {been obtained \cite{Wang:2019nes, Kelly:2019yxg, Christ:2019kaon} }. Unfortunately,
  the inclusion
  of isospin-breaking and QED effects will still take more time.
\item An independent analysis based on hadronic matrix elements from the DQCD approach \cite{Buras:2015xba, Buras:2016fys} gave a strong support
  to these values and moreover provided an \textit{upper bound} on $\epe$ in the
  ballpark of $6\times 10^{-4}$. However, this bound does not include the effects of final state interactions and it will be of interest to see how it will
  be modified when the latter are taken into account.
\item A different view has been expressed in Refs. \cite{Cirigliano:2019zjv,Cirigliano:2019ani,Gisbert:2017vvj,Cirigliano:2019cpi}, where, using
  ideas from ChPT, the authors found
  $\epe = (14 \pm 5) \times 10^{-4}$ after the improved estimate of isospin-breaking corrections to $\epe$.
  While in agreement with the measurement,
  the large uncertainty, that expresses the difficulties in matching
  long-distance and short-distance contributions in this framework, does not
  allow for any clear-cut conclusions.  See also Refs. \cite{ Buras:2016fys, Buras:2018ozh} for a critical     analysis of this approach as used in the context of $\epe$.
  \item The preliminary result on NNLO QCD corrections to QCDP contributions
  \cite{Cerda-Sevilla:2016yzo, Cerda-Sevilla:2018hjk} demonstrates
  significant reduction of various scale uncertainties, foremost of $\mu_c$,
  and indicates an additional, though modest, suppression of $\epe$.
\end{itemize}

In contrast to the expected RBC-UKQCD result, the ChPT analysis includes
isospin-breaking and QED corrections, but the known difficulties in matching
long-distance and short-distance contributions in this approach imply a large
uncertainty. In particular, the absence of the meson evolution in
ChPT that suppresses $\epe$ within the DQCD approach \cite{Buras:2016fys, Buras:2018ozh}
is responsible for the poor matching and the large value of $\epe$ quoted above.
The DQCD analysis \cite{Buras:2018lgu} demonstrates on the example of
BSM matrix elements in $K^0-\oL{K}^0$ mixing that the effects of meson evolution
are included in the present LQCD calculations. As shown in Ref. \cite{Buras:2018lgu},
neglecting this evolution in the case of $K^0-\oL{K}^0$ mixing would miss the
values of the relevant hadronic matrix by factors of $2-4$, totally misrepresenting
their values obtained by three LQCD collaborations \cite{Carrasco:2015pra,
Jang:2015sla, Garron:2016mva, Boyle:2017skn, Boyle:2017ssm}. Therefore, without
the inclusion of these important QCD dynamics in the calculation of $\epe$, the
validity of the present ChPT result can be questioned.

Now all the analyses of $\epe$  until International Conference on Kaon Physics 2019, including the  one in Ref. \cite{Cirigliano:2019cpi},
used the known Wilson coefficients at the NLO level \cite{Buras:1991jm,
Buras:1992tc, Buras:1992zv, Buras:1993dy, Ciuchini:1992tj, Ciuchini:1993vr} in
the NDR scheme \cite{Buras:1989xd}. But
already in Ref. \cite{Buras:1999st} and recently in Refs. \cite{Aebischer:2018csl, Buras:2018ozh}
it has been pointed out that without NNLO QCD corrections to EWP contribution the
results for $\epe$ are renormalization-scheme dependent and exhibit significant
non-physical dependences on the scale $\mu_t$ at which the top-quark mass
$m_t(\mu_t)$ is evaluated as well as on the matching scale $\mu_W$.

Fortunately, all these uncertainties have been significantly reduced in the NNLO
matching at the electroweak scale performed in Ref. \cite{Buras:1999st} and it is
of interest to look at them again in the context of new analyses with the goal
to improve the present estimate of $\epe$. 

Now the LQCD calculations contain the meson evolution and have recently improved on FSI effects. On the other hand ,DQCD has difficulties with the inclusion
of the latter, while ChPT has difficulties in matching long distance and short
distance contributions. Therefore, the
optimal strategy for the evaluation of $\epe$, as of 2019, appears to be  as
follows \cite{Aebischer:2019mtr}:

{\bf Step 1:} Use future RBC-UKQCD results for hadronic matrix elements of the
  dominant QCDP ($Q_6$) and EWP ($Q_8$) operators, represented by the parameters
  $\bsi$ and $\bei$ respectively -- with improved values of $\pi\pi$-strong-interaction
  phases $\delta_{0,2}$ -- but determine hadronic matrix elements
  of $(V-A)\otimes(V-A)$ operators from the experimental data on the real
  parts of the $K\to\pi\pi$ amplitudes, as done in Refs. \cite{Buras:1993dy,Buras:2015yba}.
  
{\bf Step 2:} Use the result for isospin-breaking and QED corrections from Ref.
  \cite{Cirigliano:2019cpi}, which are compatible with the ones obtained
  already 30 years ago in Ref. \cite{Buras:1987wc}.

  {\bf Step 3:}
 Use the NNLO QCD contributions to EWP in \cite{Buras:1999st} in order
 to reduce the unphysical renormalization scheme and scale dependences in the EWP sector.
 
{\bf Step 4:} Include NNLO QCD contributions to QCDP from \cite{Cerda-Sevilla:2016yzo,
  Cerda-Sevilla:2018hjk} in order to reduce left-over renormalization scale
uncertainties.

In view of the fact that meson evolution and the remaining three effects
tend to suppress $\epe$, the expectation based on the DQCD approach in Ref. 
\cite{Buras:2018ozh} that $\epe\approx (5\pm2) \times 10^{-4}$ in the SM is likely to be confirmed soon by LQCD.

The main goal of Ref. \cite{Aebischer:2019mtr} was to illustrate the importance of isospin-breaking and QED corrections \cite{Cirigliano:2019cpi}
and of the NNLO QCD contributions to EWP in Ref. \cite{Buras:1999st}, that were
absent in the 2015 result of RBC-UKQCD and also in
 Refs. \cite{Buras:2015yba,Kitahara:2016nld}.

\begin{table}
\centering
\renewcommand{\arraystretch}{1.3}
\begin{tabular}{|l|rrrr|}
\hline
             &  $a^\text{QCDP}$ & $a_6^{(1/2)}$ & $a^\text{EWP}$ & $a_8^{(3/2)}$ \\
\hline\hline
  \multicolumn{5}{|c|}{$\mu_W = \mu_t = m_W$} \\
\hline
  NLO        & $-4.19$       & $17.68$       & $-2.08$       & $8.25$        \\
  NNLO (EWP) & $-4.19$       & $17.68$       & $-2.00$       & $8.82$        \\
\hline
  \multicolumn{5}{|c|}{$\mu_W = m_W$ and  $\mu_t = m_t$} \\
\hline
  NLO        & $-4.18$       & $17.63$       & $-1.94$       & $7.22$        \\
  NNLO (EWP) & $-4.18$       & $17.63$       & $-2.03$       & $8.51$        \\
\hline
\end{tabular}
\caption{\small
  Coefficients entering the semi-numerical formula of Eq. \refeq{eq:semi-num-1}.}
  \label{tab:semi-num-1}
\end{table}

Using the technology in Ref. \cite{Buras:2015yba} the analysis in Ref. \cite{Aebischer:2019mtr}  arrives at the
formula ($a=1.017$)
\be
  \label{eq:semi-num-1}
  \boxed{\frac{\varepsilon'}{\varepsilon}  =
  \im \lambda_t \cdot \left[
   a  (1 - \OmHatEff) \left(a^\text{QCDP} + a_6^{(1/2)} \bsi \right)
    - a^\text{EWP} - a_8^{(3/2)} \bei
  \right]\,,}
\ee
with the numerical values of the coefficients given in
\reftab{tab:semi-num-1} at NLO and NNLO from EWPs as discussed below.
Explicit formulae for 
$a^\text{QCDP} = a_0^{(1/2)} - b\, a_{0,\text{EWP}}^{(1/2)}$, 
$a^\text{EWP} = a_0^{(3/2)} - a_{0,\text{EWP}}^{(1/2)}$,
$a_6^{(1/2)}$ and $a_8^{(3/2)}$ in terms of Wilson coefficients and 
$\re A_{0,2}$ are given in Ref. \cite{Buras:2015yba}, where we have 
introduced $a_{0,\text{EWP}}^{(1/2)}$ as the EWP contribution to $a_0^{(1/2)}$
and $b^{-1} = a (1 - \OmHatEff)$. Other details are presented in Ref. \cite{Aebischer:2019mtr}.
$\lambda_t = V_{td}^{} V^*_{ts}$ is the
relevant CKM combination.

The $\bsi$ and $\bei$ parameters, that enter the formula of Eq. \refeq{eq:semi-num-1},
are defined as follows
\begin{align}
  \label{eq:Q60}
  \langle Q_6(\mu) \rangle_0 &
  = -\,4 h \left[\frac{m_K^2}{m_s(\mu) + m_d(\mu)}\right]^2 (F_K - F_\pi) \,\bsi \,
  = -0.473\, h \bsi \gev^3 \,,
\\
  \label{eq:Q82}
  \langle Q_8(\mu) \rangle_2 &
  = \sqrt{2} h \left[ \frac{m_K^2}{m_s(\mu) + m_d(\mu)} \right]^2 F_\pi \,\bei\,
  = 0.862\, h \bei \gev^3 \,.
\end{align}
In the large-$N$ limit $ \bsi = \bei = 1$ \cite{Buras:1985yx,Buras:1987wc}.
We have introduced the factor $h$ in order to emphasize different normalizations
of these matrix elements present in the literature. For instance RBC-UKQCD and the authors of Ref. \cite{Buras:2015yba} use $h=\sqrt{3/2}$, while $h=1$ is used in Refs. \cite{Buras:2015xba, Buras:2016fys,Gisbert:2017vvj,Cirigliano:2019cpi} .

As an example we will first use the values  \cite{Aebischer:2019mtr} 
\begin{align}
  \label{Lbsi}
  \bsi(m_c) & = 0.80\pm 0.08 , &
  \bei(m_c) & = 0.76\pm 0.04 ,
\end{align}
to be compared with the 2015 values $\bsi(m_c) = 0.57 \pm 0.19$ and
$\bei(m_c) = 0.76 \pm 0.05$ from RBC-UKQCD \cite{Bai:2015nea, Blum:2015ywa}.
The increase of $\bsi$ could   be caused, as expected from ChPT, by an improved treatment of FSI in the update from the latter lattice collaboration.

\subsection{Scale uncertainties at NLO}
It should be emphasized that although the NLO QCD analyses of
$\epe$ in Refs. \cite{Buras:1991jm, Buras:1992tc, Buras:1992zv, Buras:1993dy,
  Ciuchini:1992tj, Ciuchini:1993vr} reduced renormalization scheme dependence in
the QCDP sector, the dependence of $\epe$ on the choice of $\mu_t$ in
$m_t(\mu_t)$ remained. This dependence can only be removed through the NNLO QCD
calculations, but in the QCDP sector it is already weak at the NLO level because
of the weak dependence of the QCDP contributions on $m_t$. On the other hand, as
pointed out already in Ref. \cite{Buras:1999st}, the EWP contributions at the NLO
level suffer from a number of unphysical dependences.
\begin{itemize}
\item First of all there is the renormalization-scheme dependence with $\epe$ in
  the HV scheme, as used in Refs. \cite{Ciuchini:1992tj, Ciuchini:1993vr}, generally
  smaller than in the NDR scheme used in Refs. \cite{Buras:1991jm, Buras:1992tc,
    Buras:1992zv, Buras:1993dy}. In what follows we will consider only the NDR
  scheme as this is the scheme used by the RBC-UKQCD collaboration and other
  analyses listed above.
\item The dependence on $\mu_t$, which is much larger than in the QCDP sector
  because the EWP contributions exhibit much stronger dependence on $m_t$
  as pointed out 30 years ago \cite{Flynn:1989iu, Buchalla:1989we}.
  Increasing $\mu_t$ makes the value of $m_t$ smaller, decreasing the EWP
  contribution and thereby making $\epe$ larger. At NLO there is no QCD
  correction that could cancel this effect.
\item The dependence on the choice of the matching scale $\mu_W$. It turns out
  that with increasing $\mu_W$ in the EWP contribution, the value of $\epe$
  decreases.
\end{itemize}

\begin{figure}
\centering
  \includegraphics[width=0.5\textwidth]{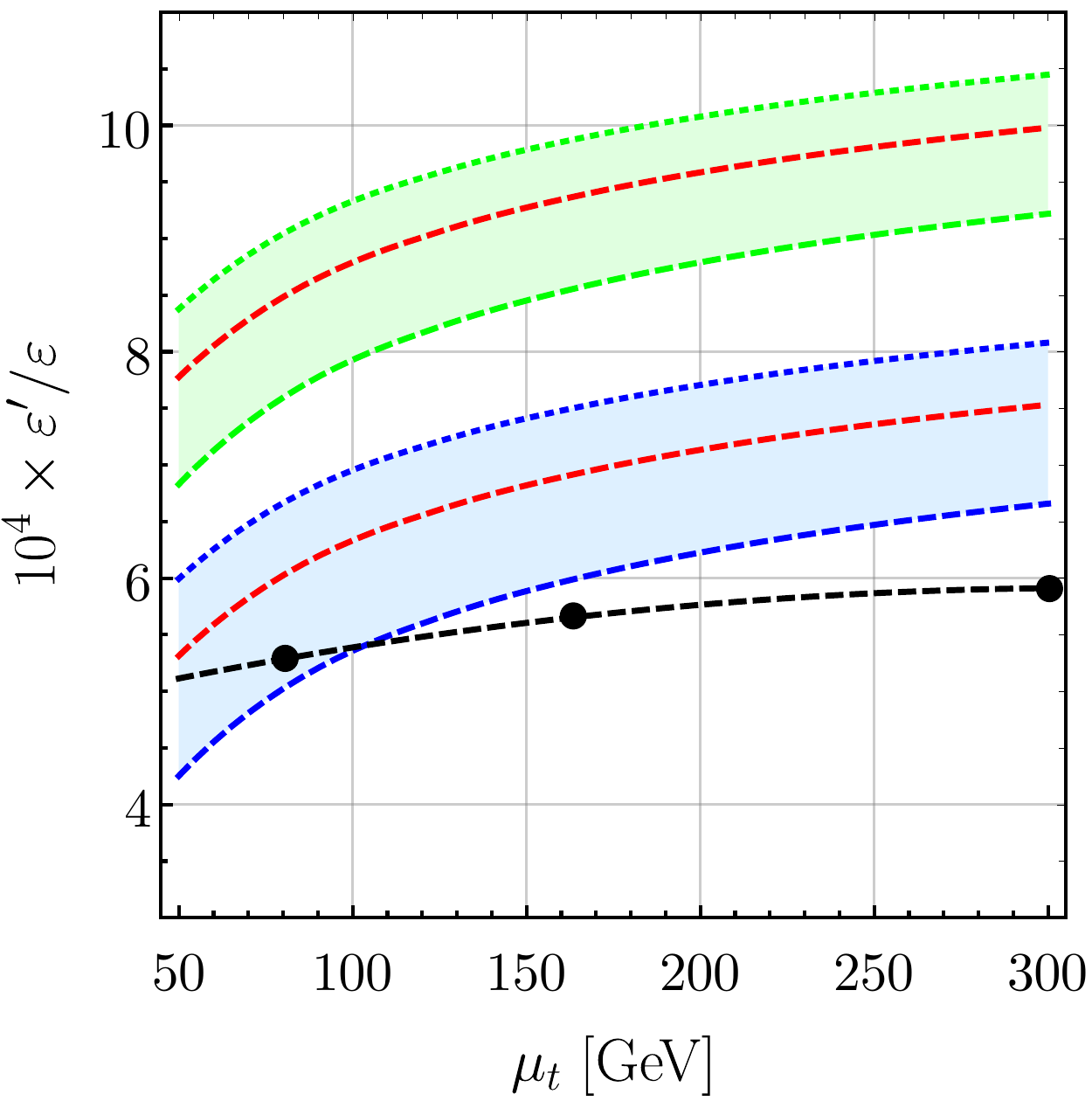}
\caption{\small The dependence of $\epe$
    for $\bsi = 0.80$ on the scale $\mu_t$  of $m_t(\mu_t)$ for three values of
    the matching scale $\mu_W = \{60,\, 80,\, 120\}\,$GeV [dotted, red, dashed]
    for $\OmHatEff = 0.0$ [green] and $\OmHatEff = 0.17$ [blue]. The black dots
    show the NNLO result for $\OmHatEff = 0.17$ at three scales $\mu_t$
    from Ref. \cite{Buras:1999st} with interpolation shown by the dashed line.
    We set $\bei=0.76$. From Ref. \cite{Aebischer:2019mtr}.}
\label{fig:epe-muW}
\end{figure}

One should note that the scales $\mu_W$ and $\mu_t$ can be chosen to be equal
or different from each other and they could be varied independently in the ranges
illustrated in \reffig{fig:epe-muW}, implying
significant uncertainties in the NLO prediction for $\epe$ as demonstrated in Ref. \cite{Buras:1999st}. In obtaining the values in \reftab{tab:semi-num-1} we
provide the two settings from Ref. \cite{Buras:1999st}: $i)$ $\mu_W = \mu_t = m_W$
 and $ii)$ $\mu_W = m_W$ and $\mu_t = m_t$. For example $ii)$ has been
used in Ref. \cite{Buras:2015yba}. Other choices of these scales, like $\mu_t=300\gev$,  would significantly
change the NLO values of $\epe$ with significantly reduced change  when NNLO
corrections to EWPs are included.

We next evaluate $\epe$ for the values of $\bsi$ and $\bei$
given in \refeq{Lbsi} and
\begin{itemize}
\item set $\mu_W = m_W$ and $\mu_t = m_t$ in the NLO formulae in the NDR scheme,
\item set isospin breaking and QED corrections to zero $\OmHatEff = 0.0$, 
\end{itemize}
as done by RBC-UKQCD. This results at NLO in
\begin{align}
  \label{RBCUKQCD}
  (\epe)_{\text{NLO},\, \OmHatEff = 0.0} &
  = (9.4 \pm 3.5) \times 10^{-4} \,.
\end{align}
The quoted error is a guess estimate based on the
uncertainties in Eq. \refeq{Lbsi} and scale uncertainties as well as the omission
of isospin-breaking effects. But as
we will see soon its precise size is irrelevant for the point we want to make.
The result in \refeq{RBCUKQCD}  is compatible  with the experimental result of Eq. \refeq{EXP}
with a tension of $1.7\,\sigma$.

At first sight it would appear that this result confirms the claims in Refs. 
\cite{Gisbert:2017vvj,Cirigliano:2019cpi,Cirigliano:2019ani} because the result of Eq. \refeq{RBCUKQCD}
is quite consistent with ChPT estimate $\epe=(14\pm 5)\cdot 10^{-4}$. But such a conclusion would be false
as we will illustrate now.

Indeed, as stated above at the NLO level, significant dependences on $\mu_W$ and
$\mu_t$ are present and the impact of a non-vanishing $\OmHatEff$ is very significant.
In order to exhibit these dependences we vary in \reffig{fig:epe-muW} the matching
scale $\mu_W$ independently of the scale $\mu_t$ at which the top-quark mass
$m_t(\mu_t)$ is evaluated and plot $\epe$ versus $\mu_t$ for the three values of
$\mu_W = \{60,\, 80,\, 120\}\,$GeV. We show these significant dependences both for
$\OmHatEff = 0.0$ [green] {and $\OmHatEff = 0.17$} [blue]. 

Fortunately all these uncertainties have been significantly reduced in the NNLO
matching at the electroweak scale performed in Ref. \cite{Buras:1999st}. In the
NDR scheme, used in all recent analyses, these corrections enhance the EWP contribution implying  a {\em negative} shift in $\epe$ as evident from \reffig{fig:epe-muW}. Including NNLO QCD corrections
in question and isospin breaking corrections from \cite{Cirigliano:2019cpi}, $\OmHatEff = 0.17\pm0.09$,
the result in Ref. \refeq{RBCUKQCD}
is changed with $\im \lambda_t = (1.43\pm0.04)\times 10^{-4}$
to \cite{Aebischer:2019mtr}
\be
  \label{ABB}
  \boxed{\epe   = (5.6 \pm 2.4) \times 10^{-4} \,.}
\ee
Compared with the experimental value in Eq. \refeq{EXP}, it signals an anomaly
at the level of $3.3\,\sigma$.
But one should keep in
mind that the central value in Eq. \refeq{ABB} will be shifted down by NNLO QCD
corrections to QCDP by about $0.5\times 10^{-4}$, as indicated in  the preliminary
plots in Refs. \cite{Cerda-Sevilla:2016yzo, Cerda-Sevilla:2018hjk} without modifying
the error in (\ref{ABB}). I am looking forward to the final results of these
authors.

\begin{table}
\centering
\renewcommand{\arraystretch}{1.3}
\begin{tabular}{|l|rrrrrrrrrrr|}
\hline
  $\OmHatEff$ $\Big/$ $\bsi$
              &       0.50 &  0.55 &  0.60 & 0.65 & 0.70 & 0.75 & 0.80 &  0.85 &  0.90 &  0.95 &  1.00 \\
\hline\hline
  \multicolumn{12}{|c|}{NLO} \\
\hline
  $0.0$ (A)   &       2.25 & 3.50  & 4.75  & 6.01 & 7.26 & 8.51 & 9.76 & 11.02 & 12.27 & 13.52 & 14.77 \\
  $0.0$ (B)   &       1.63 & 2.89  & 4.14  & 5.40 & 6.65 & 7.91 & 9.16 & 10.42 & 11.67 & 12.93 & 14.18 \\
  $0.0$ (C)   &       0.75 & 2.01  & 3.27  & 4.53 & 5.79 & 7.05 & 8.30 &  9.56 & 10.82 & 12.08 & 13.34 \\
\hline  \multicolumn{12}{|c|}{NNLO ($\mu_t = m_W$)} \\
\hline
  $0.0$       &       0.02 &  1.28 &  2.54 & 3.80 & 5.06 & 6.32 & 7.58 &  8.83 & 10.09 & 11.35 & 12.61 \\
  $0.10$      &      -0.64 &  0.50 &  1.63 & 2.76 & 3.89 & 5.03 & 6.16 &  7.29 &  8.42 &  9.56 & 10.69 \\
  $0.15$      &      -0.97 &  0.10 &  1.17 & 2.24 & 3.31 & 4.38 & 5.45 &  6.52 &  7.59 &  8.66 &  9.73 \\
  $0.20$      &      -1.30 & -0.29 &  0.71 & 1.72 & 2.73 & 3.74 & 4.74 &  5.75 &  6.76 &  7.76 &  8.77 \\
  $0.25$      &      -1.63 & -0.69 &  0.26 & 1.20 & 2.15 & 3.09 & 4.03 &  4.98 &  5.92 &  6.87 &  7.81 \\
  $0.30$      &      -1.96 & -1.08 & -0.20 & 0.68 & 1.56 & 2.44 & 3.33 &  4.21 &  5.09 &  5.97 &  6.84 \\
\hline
\end{tabular}
\caption{\small
  The ratio $10^4 \times \epe$ at NNLO for different values of the isospin
  corrections $\OmHatEff$ and the parameter $\bsi(m_c)$ with more details 
  in  \cite{Aebischer:2019mtr} and fixed value of $\bei = 0.76$ and $\im \lambda_t
  = 1.4 \times 10^{-4}$. In the first two rows we provide for comparison
  the NLO result for $\mu_t = 300\,$GeV (A), $\mu_t = m_t$ (B) and $\mu_t = m_W$ (C),
  respectively.The results for $\bsi\ge 1.0$
can be found in  Ref. \cite{Aebischer:2019mtr}.}
  \label{tab:epsp}
\end{table}

The largest remaining uncertainties in the evaluation of
$\epe$ are present in the values of $\langle Q_6(m_c)\rangle_0$ (or $\bsi$) and
$\OmHatEff$. In \reftab{tab:epsp} we give $\epe$ as a function of these two
parameters for $\bei=0.76$. This table should facilitate monitoring the values
of $\epe$ in the SM when the LQCD calculations of hadronic matrix elements,
including isospin-breaking corrections and QED effects, will improve with
time. We observe a large sensitivity of $\epe$ to $\bsi$, but for $\bsi\ge 0.7$
also the dependence on $\OmHatEff$ is significant. 
\section{Summary and Outlook}\label{sec:5}

Our analysis and in particular the comparision of the results in Eqs. \refeq{RBCUKQCD}
and \refeq{ABB}, as well as \reftab{tab:epsp}, demonstrate the importance of
NNLO QCD corrections and of isospin-breaking effects. Anticipating that the new
RBC-UKQCD analysis will find $\bsi(m_c)< 1.0$, as hinted by DQCD, the values of
$\epe$ in the SM will be significantly below the data. Our example with $\bsi(m_c)$
in the ballpark of $0.80 \pm 0.08$ illustrates a significant anomaly in $\epe$ of
{about $3.3\,\sigma$}.

However, even if $\epe$ anomaly hinted by DQCD, would be confirmed by 
new RBC-UKQCD results, it is very important to perform a number of the
following steps:
\begin{itemize}
\item Obtain satisfactory precision on $\langle Q_6(m_c)\rangle_0$ or $\bsi$.
\item Reduce the error on $\OmHatEff$. In particular isospin-breaking and QED
  effects should be taken into account in LQCD calculations.
\item Even if the insight from DQCD allowed us to identify the dynamics (meson
  evolution) responsible for this anomaly, at least a second lattice QCD
  collaboration should calculate $K\to \pi\pi$ matrix elements and $\epe$.
\item Include the NNLO QCD corrections to the QCD
  penguin sector \cite{Cerda-Sevilla:2016yzo, Cerda-Sevilla:2018hjk} and 
 the subleading NNLO QCD contributions to the electroweak
 penguin sector.
\item Calculate BSM $K\to\pi\pi$ hadronic matrix elements of four-quark
  operators by lattice QCD. They are presently only known in the DQCD
  \cite{Aebischer:2018rrz}.
\end{itemize}

\section*{Acknowledgements}
 It is  a pleasure to thank first of all Jason Aebischer and Christoph Bobeth
 for the collaboration in the context of Ref. \cite{Aebischer:2019mtr}. Particular thanks go to Jean-Marc G{\'e}rard for our intensive work on DQCD and to
 Jason Aebischer, Christoph Bobeth and  David  Straub for the study of $\epe$
 in the context of the SMEFT.
I would also like to thank the organizers of International Conference on Kaon Physics 2019  for inviting me 
to this very interesting conference  and for an impressive hospitality.
This research was supported by the Excellence Cluster ORIGINS,
funded by the Deutsche Forschungsgemeinschaft (DFG, German Research Foundation) under Germany´s Excellence Strategy – EXC-2094 – 390783311.

\section*{References}

\bibliographystyle{iopart-num}
\bibliography{Book19}
\end{document}